\documentclass[12pt]{article}
\usepackage{color}
\usepackage{amssymb,amsmath,comment}
\usepackage{graphicx}
\usepackage{braket}
\usepackage[dvipdfmx]{hyperref}
\usepackage{epsfig}
\usepackage{here}
\graphicspath{{./Figures/}}
\newcommand{\ba}{\begin{align}}
\newcommand{\ea}{\end{align}}

\def\nn{\nonumber}

\def\bea{\begin{eqnarray}}
\def\eea{\end{eqnarray}}
 \makeatletter
\def\alt{\mathrel{\mathpalette\gl@align<}}
\def\agt{\mathrel{\mathpalette\gl@align>}}
\def\gl@align#1#2{\lower.6ex\vbox{\baselineskip\z@skip\lineskip\z@
\ialign{$\m@th#1\hfil##\hfil$\crcr#2\crcr\sim\crcr}}} \makeatother
\renewcommand{\thefootnote}{\fnsymbol{footnote}}
\newcommand{\doi}[1]{\href{https://doi.org/#1}{doi:#1}}
\begin{document}
\begin{flushright}
\end{flushright}
\vspace*{1.0cm}

\begin{center}
\baselineskip 20pt 
{\Large\bf 
Proton Lifetime Upper Bound \\
in Non-SUSY SU(5) GUT
}
\vspace{1cm}

{\large 
Naoyuki Haba,
\ Yukihiro Mimura
\ and \ Toshifumi Yamada
} \vspace{.5cm}

{\baselineskip 20pt \it
Institute of Science and Engineering, Shimane University, Matsue 690-8504, Japan
}

\vspace{.5cm}

\vspace{1.5cm} {\bf Abstract} \end{center}

In preparation for upcoming nucleon decay searches at Hyper-Kamiokande,
 it is important to derive a theoretical upper bound on the proton lifetime
 in a general class of grand unified theory (GUT) models.
In this paper, we make an attempt along this direction for non-SUSY SU(5) models, under the mild restrictions that 
 only one or two SM-decomposed multiplets are singularly light, and that the SU(5) gauge theory is asymptotically free and thus there are no too large representations in the model.
We derive criteria for SM-decomposed multiplets that potentially enhance the proton lifetime when they are singularly light.
We perform a numerical analysis on the proton lifetime and show that 
some choices of singularly light multiplets can provide
 a testable upper bound on the proton lifetime.

\thispagestyle{empty}

\newpage
\renewcommand{\thefootnote}{\arabic{footnote}}
\setcounter{footnote}{0}
\baselineskip 18pt
\section{Introduction}

The Hyper-Kamiokande (HK) experiment is expected to start operation in 2026,
 and after a 10~year exposure of one 187~kton fiducial volume detector,
 HK can make a 3$\sigma$ discovery of $p\to e^+\pi^0$ process
 for the partial lifetime up to $\tau_p=6.3\times 10^{34}$~years~\cite{Abe:2018uyc}.
Now that the time schedule of nucleon decay searches is settled,
 it is time to inspect which types of grand unified theory (GUT) models~(for a review, see \cite{Langacker:1980js})
 are possibly discovered by HK.
For this purpose, it is important to derive a \textit{theoretical upper bound} of the proton lifetime
 in a general class of models, rather than elaborate a specific model to prolong the proton lifetime.

In the minimal supersymmetric (SUSY) SU(5) GUT~\cite{Dimopoulos:1981zb,Dimopoulos:1981yj,Sakai:1981gr},
 it is absolutely impossible to derive a theoretical upper bound on the proton lifetime.
For processes mediated by the GUT gauge boson (dimension-six processes), the reason is as follows:
The gauge coupling unification conditions give
\bea
M_X^4M_{({\bf8},{\bf1},0)}M_{({\bf1},{\bf3},0)}\simeq (2\times 10^{16}~{\rm GeV})^6,
\label{gcu}
\eea
 where $M_X$ denotes the GUT gauge boson mass, and $M_{({\bf8},{\bf1},0)}$ and $M_{({\bf1},{\bf3},0)}$ represent 
 the masses of ${({\bf8},{\bf1},0)}$ and ${({\bf1},{\bf3},0)}$ multiplets in the GUT breaking Higgs representation decomposed under the Standard Model (SM) gauge groups.
Eq.~(\ref{gcu}) tells us that $M_X$ is inversely proportional to $(M_{({\bf8},{\bf1},0)}M_{({\bf1},{\bf3},0)})^{1/4}$.
The issue here to derive the bound
is that $M_{({\bf8},{\bf1},0)}$,\,$M_{({\bf1},{\bf3},0)}$ are proportional to the self-coupling of the GUT breaking Higgs representation.
Since there is no theoretical lower bound on this self-coupling, $M_{({\bf8},{\bf1},0)}$ and $M_{({\bf1},{\bf3},0)}$ can be lowered arbitrarily.
Consequently, $M_X$ can be enhanced arbitrarily, and hence no upper bound on the proton lifetime in dimension-six processes
 \footnote{
 By decreasing $M_{(8,1,0)}$ and $M_{(1,3,0)}$ by $1/10$, $M_X$ is enhanced by $\sqrt{10}$ and the proton lifetime becomes 100~times larger.
 Thus, the proton lifetime can easily exceed the HK discovery reach.
}.
For processes mediated by the colored Higgsino (dimension-five processes),
 the unification conditions can pin down the colored Higgs mass $M_{H_C}$ 
 if $M_{({\bf8},{\bf1},0)}$ and $M_{({\bf1},{\bf3},0)}$ are degenerate.
However, $M_{H_C}$ can be easily enlarged if $M_{({\bf8},{\bf1},0)}<M_{({\bf1},{\bf3},0)}$,
 which is possible by adding non-renormalizable terms \cite{Bajc:2002pg},
 or if there exists a non-renormalizable coupling of the GUT breaking Higgs with the SU(5) gauge kinetic term
 that modifies the unification conditions~\cite{Tobe:2003yj}.
Moreover, the proton lifetime in dimension-five processes is subject to uncertainties of
 the details of the Yukawa unification and the mass of SUSY particles.
Therefore, it is again impossible to derive an upper bound.

What about non-SUSY SU(5) GUT?
In the minimal model with only {\bf 24}-dimensional GUT breaking Higgs representation, the gauge coupling unification conditions give
\bea
M_X^{42}M_{({\bf8},{\bf1},0)}M_{({\bf1},{\bf3},0)}\simeq (5\times10^{13}~{\rm GeV})^{44}.
\label{gcu2}
\eea
Because of the big exponent $42$, $M_X$ cannot be enhanced efficiently;
 for $M_{({\bf8},{\bf1},0)},M_{({\bf1},{\bf3},0)}>1$~TeV, one gets $M_X\lesssim 2\times 10^{14}$~GeV, 
 which is below the bound at Super-Kamiokande (SK)~\cite{Miura:2016krn}.
However, if a model contains extra scalar fields other than {\bf 5} and {\bf 24}
 and if the scalar potential is tuned in such a way that one or two SM-decomposed multiplets are singularly light compared to the other multiplets in the same SU(5) representations,
 the light multiplets modify the renormalization group equations (RGEs) and possibly enhance $M_X$ above the current experimental bound
 \footnote{
 For a recent related study, see Ref~\cite{Schwichtenberg:2018cka}.
 }.
(To have a singularly light SM-decomposed multiplet(s), a fine-tuning of the scalar potential is mandatory.
In this paper, we perform a phenomenological study and do not discuss the origin of this fine-tuning.)

The next questions is, then, which choice of the singularly light SM-decomposed multiplets leads to an arbitrarily enhanced proton lifetime,
 and which choice leads to a proton lifetime bounded from above, and can further give a testable (i.e., within the scope of HK) proton lifetime upper bound.
In this paper, we answer to this question by 
 performing a systematic survey on a broad range of non-SUSY SU(5) GUT models.
We only mildly restrict our study to the cases satisfying two conditions below:
\begin{itemize}

\item
There are only one or two singularly light SM-decomposed multiplets, and the rest of the SM-decomposed multiplets are mass-degenerate with the GUT gauge boson.

\item
The SU(5) gauge theory is asymptotically free even if all the scalar particles participate in the renormalization group evolutions.
Thus, SU(5) representations with a large Dynkin index are not considered.

\end{itemize}
We will demonstrate in the main body of the paper that
 it is indeed possible to put a testable upper bound on the proton lifetime
 for some choices of singularly light SM-decomposed multiplets.

This paper is organized as follows:
In Section~2, we revisit the 1-loop gauge coupling unification conditions in non-SUSY SU(5) GUT,
 and survey SM-decomposed multiplets that potentially enhance the proton lifetime when they are singularly light.
Section~3 presents our main result;
 we display the proton lifetime for all patterns of singularly light SM-decomposed multiplets, 
 and study in which cases the proton lifetime is bounded from above.
Section~4 summarizes the paper.
\\

\section{Singularly Light SM-decomposed Multiplets that enhance the Proton Lifetime}

The gauge coupling unification conditions in non-SUSY SU(5) GUT read
\begin{eqnarray}
&&M_{H_C} \prod_i M_i^{l_A^i} \simeq 10^{87} \ {\rm GeV} \label{gcu3}\\
&&M_X^{42}  \prod_i M_i^{l_B^i} \simeq (5 \times 10^{13} \ {\rm GeV})^{44},
\label{gcu4}
\end{eqnarray}
 where $M_{H_C}$ denotes the colored Higgs mass, and $M_X$ denotes the GUT gauge boson mass.
$i$ labels a SM-decomposed multiplet other than the would-be-Nambu-Goldstone mode (i.e., other than ({\bf 3},{\bf 2},$-5/6$) of {\bf 24}).
$M_i$ denotes the mass of multiplet $i$, and $l_A^i,l_B^i$ are indices for multiplet $i$ whose values are tabulated in Tables~\ref{Tab1},\ref{Tab2}.
The derivation of Eqs.~(\ref{gcu3}),(\ref{gcu4}) is found in Appendix.

If only one SM-decomposed multiplet is singularly light compared to the other SM-decomposed multiplets belonging to the same SU(5) representation,
 Eqs.~(\ref{gcu3}),(\ref{gcu4}) are recast into
\begin{eqnarray}
&&M_{H_C} \simeq 10^{87} \ {\rm GeV} \times \left(\frac{M_r}{M_M}\right)^{-l_A^r},
\label{gcu-1p-1}\\
&&M_X  \simeq 5\times 10^{13} \ {\rm GeV} \times \left(\frac{M_X}{M_{\Sigma}}\right)^{\frac{1}{22}} \left(\frac{M_M}{M_r}\right)^{\frac{l_B^r}{44}},
\label{gcu-1p-2}
\end{eqnarray}
where $r$ labels the light SM-decomposed multiplet, and $M_M$ is the common mass of the other SM-decomposed multiplets in the same SU(5) representation.
$M_\Sigma$ is the mass of the physical particles in the SU(5) representation whose vacuum expectation value (VEV) breaks SU(5) and whose $({\bf3},{\bf 2},-5/6)$ component is absorbed into the GUT gauge boson.
Eliminating $M_M$ from Eqs.~(\ref{gcu-1p-1}),(\ref{gcu-1p-2}), one obtains
\begin{equation}
M_X \simeq 5 \times 10^{13} \ {\rm GeV} \times \left(\frac{M_X}{M_\Sigma}\right)^{\frac1{22}}
\left(\frac{10^{87}\  {\rm GeV}}{M_{H_c}}\right)^{-\frac{l_B^r}{44 l_A^r}}.
\label{gcu-final}
\end{equation}
If instead two SM-decomposed multiplets (which may not belong to the same SU(5) representation) are singularly light,
 Eqs.~(\ref{gcu3}),(\ref{gcu4}) can be rewritten as
\begin{eqnarray}
&&M_{H_C} \simeq 10^{87} \ {\rm GeV} \times \left(\frac{M_{\rm eff}}{M_M}\right)^{-l_A^{\rm eff}},
\label{gcu-2p-1}\\
&&M_X  \simeq 5\times 10^{13} \ {\rm GeV} \times \left(\frac{M_X}{M_{\Sigma}}\right)^{\frac{1}{22}} \left(\frac{M_M}{M_{\rm eff}}\right)^{\frac{l_B^{\rm eff}}{44}},
\label{gcu-2p-2}
\end{eqnarray}
 where $M_{\rm eff}$, $l_A^{\rm eff}$, $l_B^{\rm eff}$ are defined from the mass $M_{r_i}$ and $l_A^{r_i},l_B^{r_i}$ of the two light SM-decomposed multiplets labelled by $r_1,r_2$ as
\bea
&&\left(\frac{M_{\rm eff}}{M_M}\right)^{-l_A^{\rm eff}}=\left(\frac{M_{r_1}}{M_{M_1}}\right)^{-l_A^{r_1}}\left(\frac{M_{r_2}}{M_{M_2}}\right)^{-l_A^{r_2}},\label{leff1}\\
&&\left(\frac{M_{\rm eff}}{M_M}\right)^{-l_B^{\rm eff}}=\left(\frac{M_{r_1}}{M_{M_1}}\right)^{-l_B^{r_1}}\left(\frac{M_{r_2}}{M_{M_2}}\right)^{-l_B^{r_2}},\label{leff2}
\eea
 with $M_{M_i}$ $(i=1,2)$ being the common mass of the other SM-decomposed multiplets in the same SU(5) representation as the light multiplet $r_i$.
After eliminating $M_M$, one obtains an analogous formula as Eq.~(\ref{gcu-final}).
\\

It is interesting to compare the above conditions with the SUSY case.
The GUT gauge boson mass in SUSY SU(5) reads
\begin{equation}
M_X \simeq 2\times 10^{16} \ {\rm GeV} \times \left(\frac{M_X}{M_\Sigma}\right)^{\frac1{3}} \left(\frac{M_M}{M_r}\right)^{\frac{l_B^r}{6}}.
\end{equation}
The ratio $M_X/M_\Sigma$ is proportional to $1/\lambda$, where $\lambda$ is the self-coupling
 of the GUT Higgs field and is not bounded from below theoretically.
One finds that the proton lifetime in dimension-six processes becomes $2^{4/3}\simeq 2.5$ times larger
 when $M_X/M_\Sigma$ increases by twice (the self-coupling of the GUT Higgs field decreases by half).
Besides, both the dimension-five and six proton decay amplitudes are suppressed
if a SM-decomposed multiplet whose $l_A$ and $l_B$ are both positive is singularly light \cite{Dutta:2007ai}.
In non-SUSY case, on the other hand, 
the GUT gauge boson mass is much insensitive to the threshold corrections of scalar multiplets near the GUT scale. 
The proton lifetime becomes merely 50\% larger even if the mass ratio $M_X/M_\Sigma$ is 10 times larger,
 which is a better situation for putting an upper bound on the proton lifetime.
\\

The current bound on the dimension-six proton decay $p \to \pi^0 e^+$
 corresponds to $M_X \agt 6 \times 10^{15}$~GeV for the unified gauge coupling $\alpha_U \simeq 1/35$.
Eqs.~(\ref{gcu-1p-1}),(\ref{gcu-1p-2}) or Eqs.~(\ref{gcu-2p-1}),(\ref{gcu-2p-2}) tell us that 
 to satisfy the above bound on $M_X$ while having $M_{H_C}$ in a reasonable range below the Planck scale,
 we need a singularly light multiplet with $l_A^r < 0$ and $l_B^r>0$ and large $-l_A^r$ and $l_B^r$,
 or two multiplets with $l_A^{\rm eff}< 0$ and $l_B^{\rm eff}>0$ and large $-l_A^{\rm eff}$ and $l_B^{\rm eff}$.

If there is only one singularly light multiplet,
 one finds from Table~\ref{Tab2} two candidates for it,
\footnote{
It is understood that complex-conjugate fields are included implicitly.}
\bea
&&({\bf 6},{\bf 3} ,1/3) \ {\rm in} \ \overline{\bf 50},\\
{\rm or}&&({\bf 8},{\bf 3},0) \ {\rm in} \ {\bf 75},
\eea
which respectively yield
\bea
&&(l_A^r,l_B^r)=(-\frac{33}2, 15) \quad {\rm for}\ ({\bf 6},{\bf 3} ,1/3), \\
&&(l_A^r,l_B^r)=(-\frac{25}2, 11) \quad {\rm for}\ ({\bf 8},{\bf 3} ,0).
\eea
However, Eq.(\ref{gcu-final}) tells us that for small $-l_B^r/l_A^r$, the mass of the colored Higgs boson $M_{H_C}$ is considerably reduced when $M_X$ is enhanced.
On the other hand, $M_{H_C}$ must be larger than roughly $10^{10}$~GeV to avoid a dangerous dimension-six proton decay via the colored Higgs boson exchange,
 although the precise bound depends on the suppression from the Yukawa couplings.
From Eq.(\ref{gcu-final}), we find that $M_X \agt 6 \times 10^{15}$~GeV and $M_{H_C}\gtrsim10^{10}$~GeV are simultaneously achieved for $-l_B^r/l_A^r \agt 1.1$, 
 which is not satisfied by either candidate
\footnote{
If we drop the restriction that the SU(5) gauge theory is asymptotically free 
 and instead adopt the criterion that the SU(5) gauge coupling remains perturbative up to about $10^{18}$~GeV scale,
 we can have a larger SU(5) multiplet.
It is then possible to construct a viable GUT model with \textit{only one} singularly light SM-decomposed multiplet.
Specifically, we are allowed to introduce a ${\bf 126^\prime}(5000)$, ${\bf 175^\prime}(1200)$, ${\bf 175^{\prime\prime}}(0300)$ or ${\bf 280}(1110)$ multiplet
 without conflicting the criterion that the SU(5) gauge coupling remains perturbative up to $\sim10^{18}$~GeV,
 and if $({\bf 10},{\bf3},0)$ in ${\bf 126^\prime}(5000)$, ${\bf 175^\prime}(1200)$ or
$({\bf 15},{\bf 3},1/3)$ in ${\bf 175^{\prime\prime}}(0300)$, ${\bf 280}(1110)$
 is singularly light,
 the model satisfies $-l_B^r/l_A^r \agt 1.1$, $l_A^r < 0$, $l_B^r>0$, and the condition that $-l_A^r$ and $l_B^r$ be large.
}.

It follows that we need (at least) two singularly light multiplets, with $l_A^{\rm eff}< 0$ and $l_B^{\rm eff}>0$, $-l_A^{\rm eff}$ and $l_B^{\rm eff}$ being large, 
 and $-l_B^{\rm eff}/l_A^{\rm eff} \agt 1.1$.
There are two possible scenarios below:
\begin{enumerate}

\item
$({\bf 6},{\bf 3} ,1/3)$ or $({\bf 8},{\bf 3},0)$ is light,
and another `assisting' multiplet whose $l_A$ is positive and whose $-l_B$ is not large (positive $l_B$ is favored) makes
$-l_B^{\rm eff}/l_A^{\rm eff}$ larger.
The candidates of the assisting multiplet are
\begin{eqnarray*}
&&({\bf15},{\bf 1},-1/3),\ ({\bf 10},{\bf1},-1), \ ({\bf8},{\bf 2},1/2),\ ({\bf 8},{\bf1},1), \\
&&({\bf 6},{\bf 1},1/3),\ ({\bf 6},{\bf 1}, -2/3), \ ({\bf 6},{\bf 2},-1/6), \ \ \ {\rm etc.}
\end{eqnarray*}
The multiplets $({\bf15},{\bf 1},-1/3)$ in $\bf 70$ and $({\bf 10},{\bf1},-1)$ in $\overline{\bf 35}$
 have large $l_A$, leading to larger $M_X$ than the case when only $({\bf 6},{\bf 3} ,1/3)$  or $({\bf 8},{\bf 3},0)$ is light.
The multiplets $({\bf 6},{\bf 1},1/3)$ and $({\bf8},{\bf 2},- 1/2)$
are contained in $\overline{\bf 45}$ and they can couple to bi-fermions as ${\bf 10}_F \cdot \bar{\bf5}_F \cdot \overline{\bf45}$.
The multiplet $({\bf 6},{\bf 1},1/3)$ is identified with a di-quark,
and if it is light, it can cause $\Delta B = 2$ processes like neutron-antineutron oscillations~\cite{Babu:2012vc}.

\item
A multiplet with $l_A>0$ and $l_B>0$, such as $({\bf 6},{\bf 2},-1/6)$ and
$({\bf 8},{\bf2},1/2)$, is light, and another assisting multiplet with $l_A<0$ and $l_B>0$ allows $l_B^{\rm eff},l_A^{\rm eff}$ to satisfy the conditions.
Excluding the pairs in the scenario~1, we find the candidates of the assisting multiplet to be
\begin{equation}
({\bf3},{\bf3},-1/3),\ ({\bf 3},{\bf3},2/3),\ ({\bf1},{\bf 4},1/2).
\end{equation}
We note that $({\bf3},{\bf3},-1/3)$ is contained in $\overline{\bf 45}$ and can cause proton decay.
This multiplet should be heavier than roughly $10^{10}$ GeV, depending on the Yukawa couplings of $\overline{\bf45}$ representation.

\end{enumerate}

In the next section, we will calculate the proton lifetime along the two scenarios above.
\\

\section{Numerical Results for Proton Lifetime}

We calculate the proton lifetime for the dimension-six process in non-SUSY non-minimal SU(5) GUT using two-loop RGEs.
We exclusively consider those GUT models that satisfy two restrictions below:

\begin{itemize}

\item
There are only one or two singularly light SM-decomposed multiplets, and the rest of the SM-decomposed multiplets are mass-degenerate with the GUT gauge boson.
We have revealed in the previous section that the case with only one singularly light multiplet is not viable. Hence, we concentrate on the case with two singularly light multiplets.

\item
The SU(5) gauge theory is asymptotically free even if all the scalar particles participate in the renormalization group evolutions.
 Thus, SU(5) representations with a large Dynkin index are not considered.

\end{itemize}
Two comments are in order:

\begin{itemize}

\item
Because of the big exponent 42 for $M_X$ in Eq.~(\ref{gcu4}), the detailed scalar mass spectrum does not significantly change the proton lifetime.
Therefore, it is justifiable to ignore how the scalar mass spectrum is derived from a concrete scalar potential and approximate that all the multiplets other than the singularly light ones have the same mass
 as the GUT gauge boson.
 
\item
From the restriction that the SU(5) gauge theory is asymptotically free, the total of $l_i$ (found in Tables~\ref{Tab1},\ref{Tab2})
 must be less than 80, if we break SU(5) by a real $\bf 24$ scalar.
Accordingly, we do not employ $\bf 70'$ or any representation with 100 or higher dimensions.
Also, since the sum of $l_i$ of $\bf 50$ and $\bf 70$ is more than 80,
 we do not employ $\bf 50$ and $\bf 70$ simultaneously.

 \end{itemize}

In the calculation of the proton lifetime,
 we use the proton decay hadronic matrix element $\alpha_H = -0.014 \ {\rm GeV}^3$ at 2 GeV~\cite{Aoki:2017puj},
and chiral Lagrangian parameters $F=0.463$, $D=0.804$.
The 5-flavor strong gauge coupling at $\mu= M_Z$ is
$\alpha^{(5)}_s (M_Z) = 0.1181 \pm 0.0011$~\cite{Tanabashi:2018oca},
and we present the lifetime obtained from both $+ 3\sigma$ value and $- 3\sigma$ value of $\alpha^{(5)}_s (M_Z)$.
The proton lifetime is about factor 4 different for those two values of $\alpha^{(5)}_s (M_Z)$.
We calculate the unified gauge coupling by using two-loop RGEs~\cite{Jones:1981we,Machacek:1983tz}.
We remark that
the two-loop RG running of SU(2) gauge coupling
receives contributions from SU(3) gauge coupling if
$({\bf 6},{\bf 3},1/3)$ or $({\bf 8},{\bf 3},0)$ is light,
 which makes the GUT gauge boson mass roughly 1/2 compared to the one-loop calculation.
Consequently, the proton lifetime is one digit smaller than the one-loop results.

We fix the colored Higgs mass $M_{H_C}$ at $10^{10}$ GeV when solving the unification conditions, which gives a larger value for the proton lifetime
 than the cases when $M_{H_C}$ is larger. 
In fact, the proton lifetime is not sensitive to the colored Higgs mass, and if we take it to be $10^{16}$ GeV, the proton lifetime becomes $1/2$-$1/3$ of the values in the plots.

We show the proton lifetime as a function of the mass of the lightest SM-decomposed multiplet.
In each figure, the horizontal solid line is the current experimental bound on $p \to \pi^0 e^+$ partial lifetime~\cite{Miura:2016krn}
\begin{equation}
\tau_p > 1.6 \times 10^{34} \ {\rm years},
\end{equation}
and the horizontal dashed line is the $3\sigma$ discovery potential at HK with a 10~year exposure of 1-tank,
$6.3 \times 10^{34}$~years~\cite{Abe:2018uyc}.
The mass of the second lightest SM-decomposed multiplet also changes along each slope, and of course it is always larger than the mass of the leading lightest one.

\begin{figure}[tbp]
\includegraphics[width=8cm]{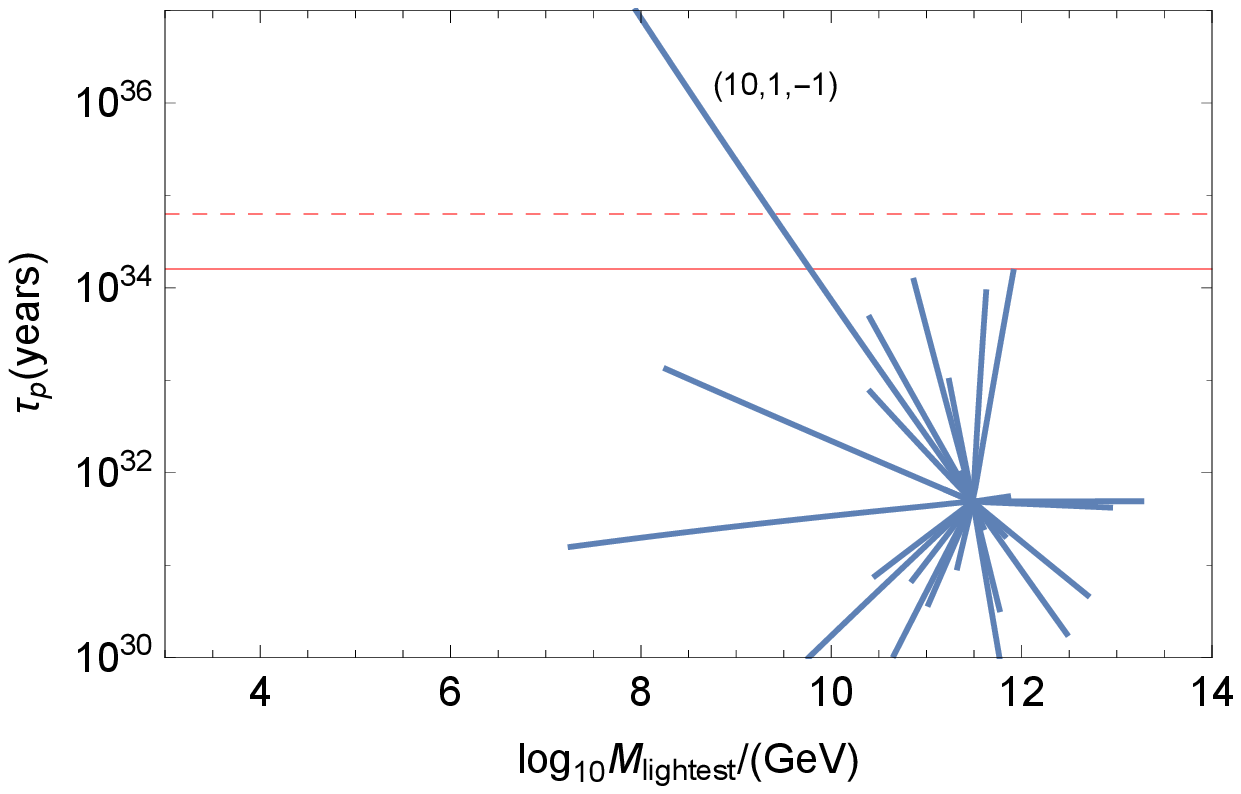}
\includegraphics[width=8cm]{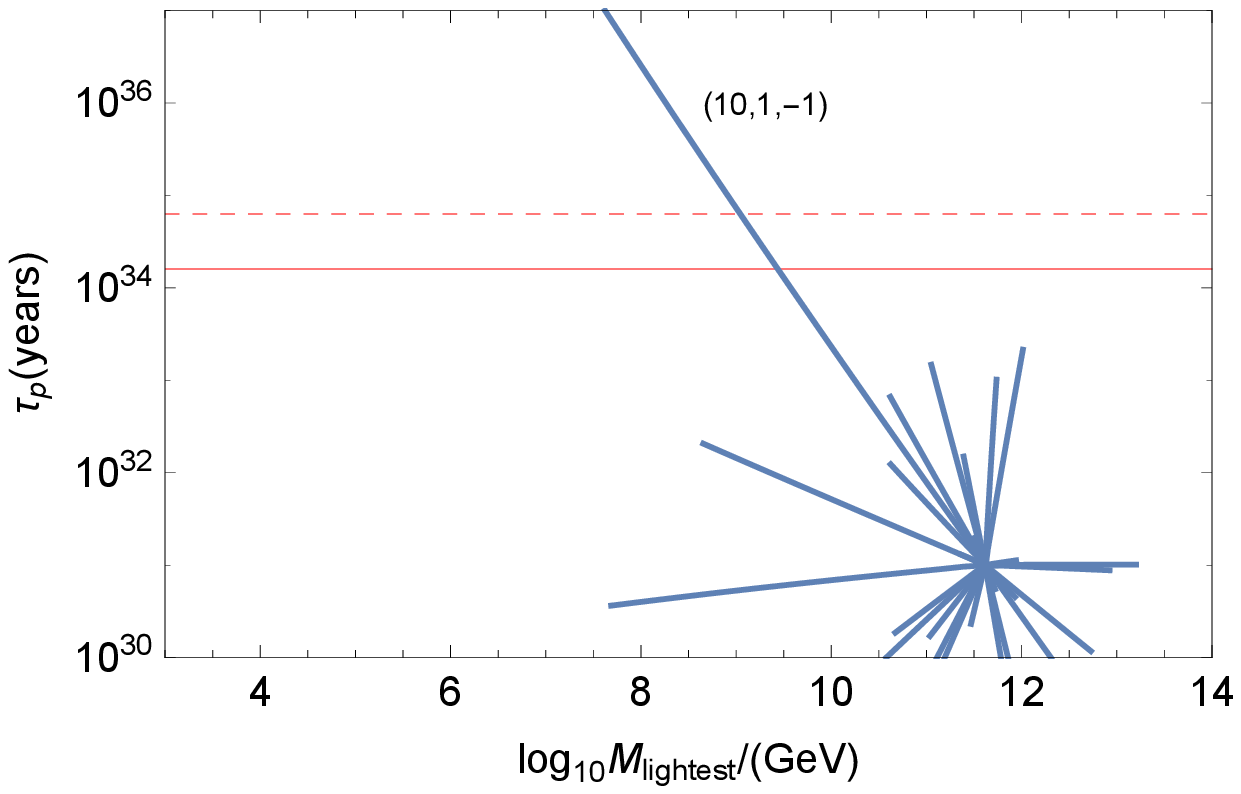}
\caption{
Proton partial lifetime for $p \to \pi^0 e^+$ process, $\tau_p$, in the case when
$({\bf 6},{\bf 3},1/3)$ is the \textit{leading} lightest SM-decomposed multiplet.
The horizontal axis is the mass of the leading lightest multiplet, which is $({\bf 6},{\bf 3},1/3)$ in this figure.
$\alpha_s^{(5)}(M_Z) = 0.1181+3\cdot0.0011$ (left), $0.1181-3\cdot0.0011$ (right).
The red lines show the current bound at SK (solid), and the $3\sigma$ discovery potential at HK with a 10~year exposure of 1-tank (dashed).
The blue line labelled as $({\bf 10},{\bf1},-1)$ shows the proton partial lifetime when $({\bf 10},{\bf1},-1)$ is the \textit{second} lightest multiplet.
The unlabelled blue lines correspond to cases with a different second lightest multiplet, and it is evident that all these cases are already excluded by SK.
}
\label{fig1}
\end{figure}

\begin{figure}[tbp]
\includegraphics[width=8cm]{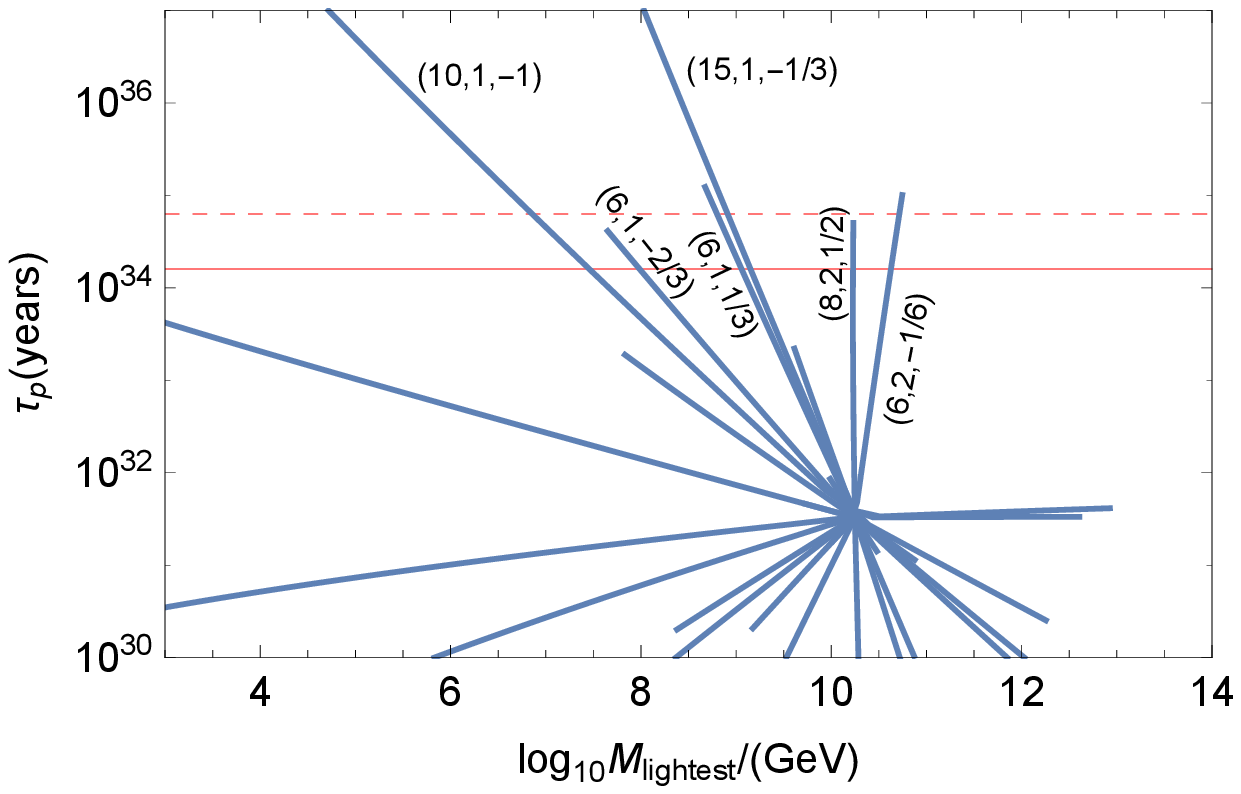}
\includegraphics[width=8cm]{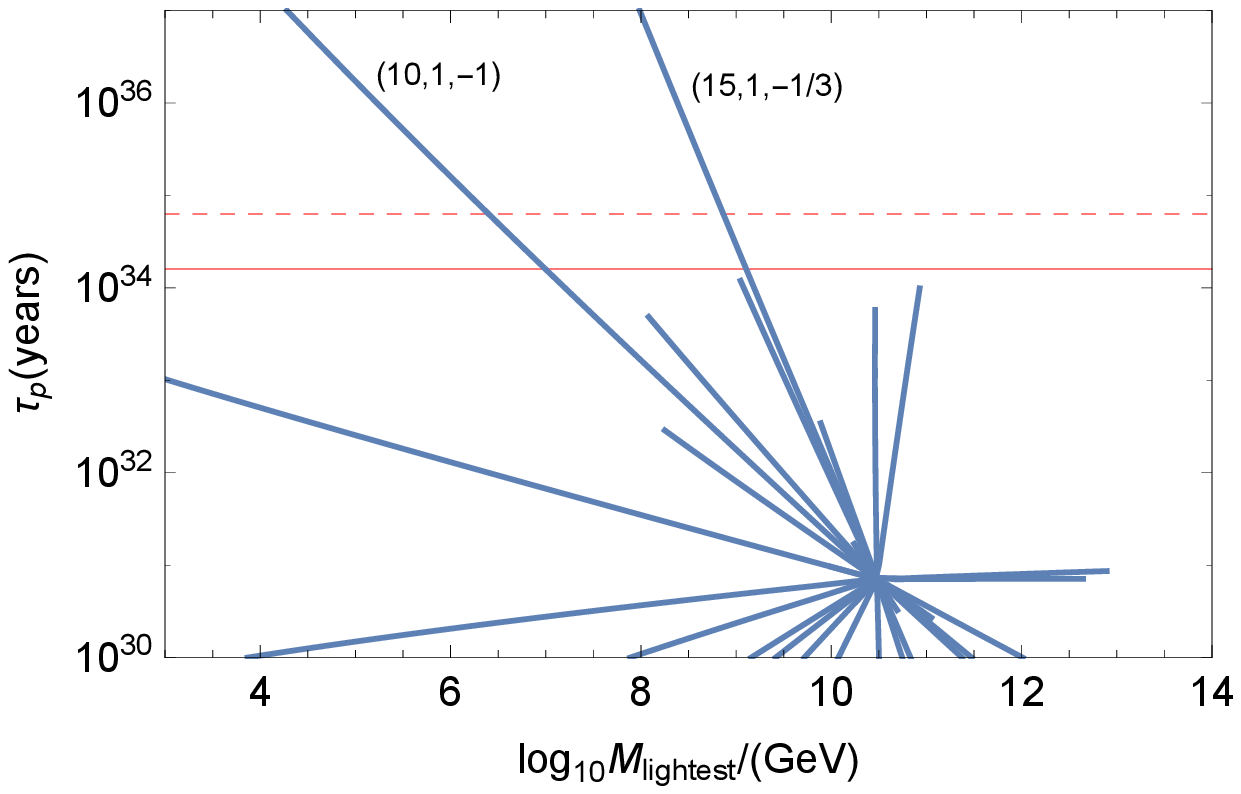}
\caption{
Proton partial lifetime, $\tau_p$, in the case when
$({\bf 8},{\bf 3},0)$ is the \textit{leading} lightest SM-decomposed multiplet.
The horizontal axis is the mass of the leading lightest multiplet, which is $({\bf 8},{\bf 3},0)$ in this figure.
$\alpha_s^{(5)}(M_Z) = 0.1181+3\cdot0.0011$ (left), $0.1181-3\cdot0.0011$ (right).
The red lines show the current bound at SK (solid), and the $3\sigma$ discovery potential at HK with a 10~year exposure of 1-tank (dashed).
Each blue line with a label shows the proton partial lifetime when the \textit{second} lightest multiplet is as indicated by the label.
The unlabelled blue lines correspond to cases with a different second lightest multiplet, and it is evident that all these cases are already excluded by SK.
Interestingly, if the second lightest multiplet is $({\bf6},{\bf 1},-2/3)$, $({\bf6},{\bf 1},1/3)$ or $({\bf 6},{\bf 2},-1/6)$, the model can evade the current bound
 and predicts a proton lifetime within the coverage of HK.
}
\label{fig2}
\end{figure}

\begin{figure}[tbp]
\includegraphics[width=8cm]{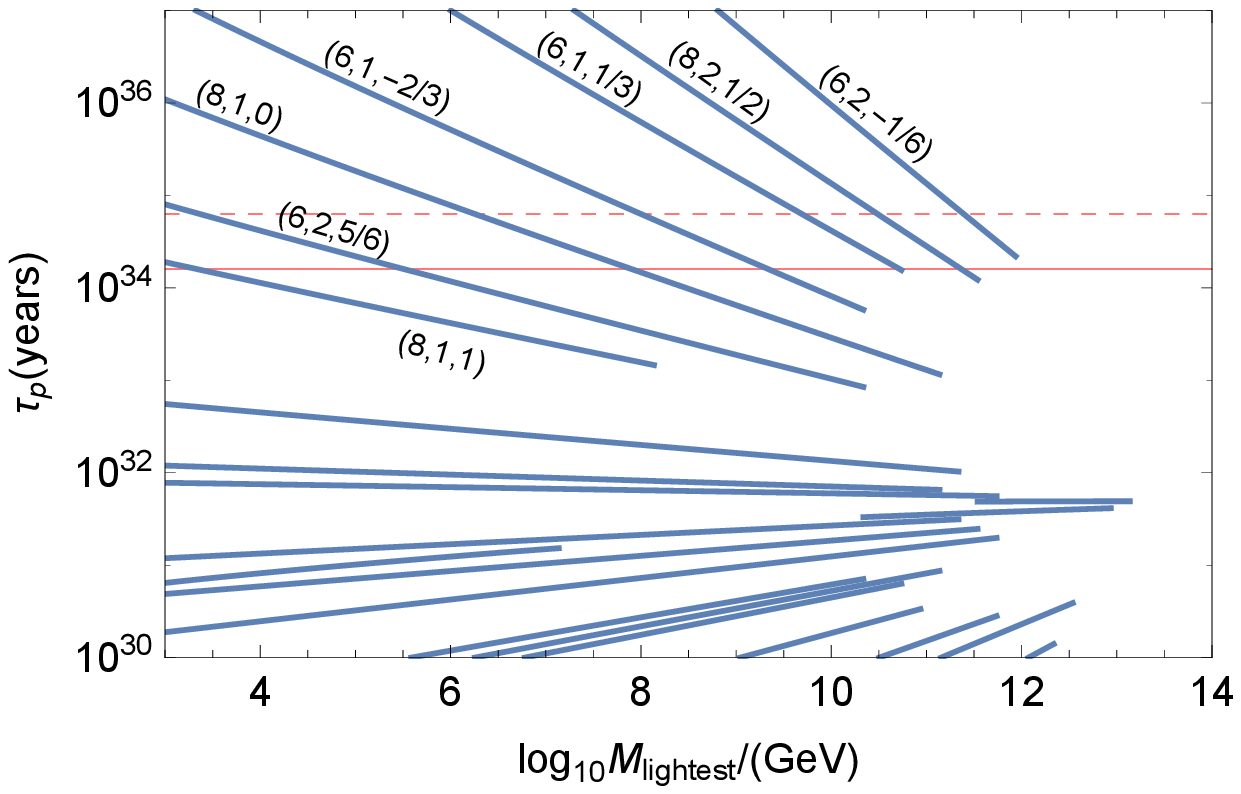}
\includegraphics[width=8cm]{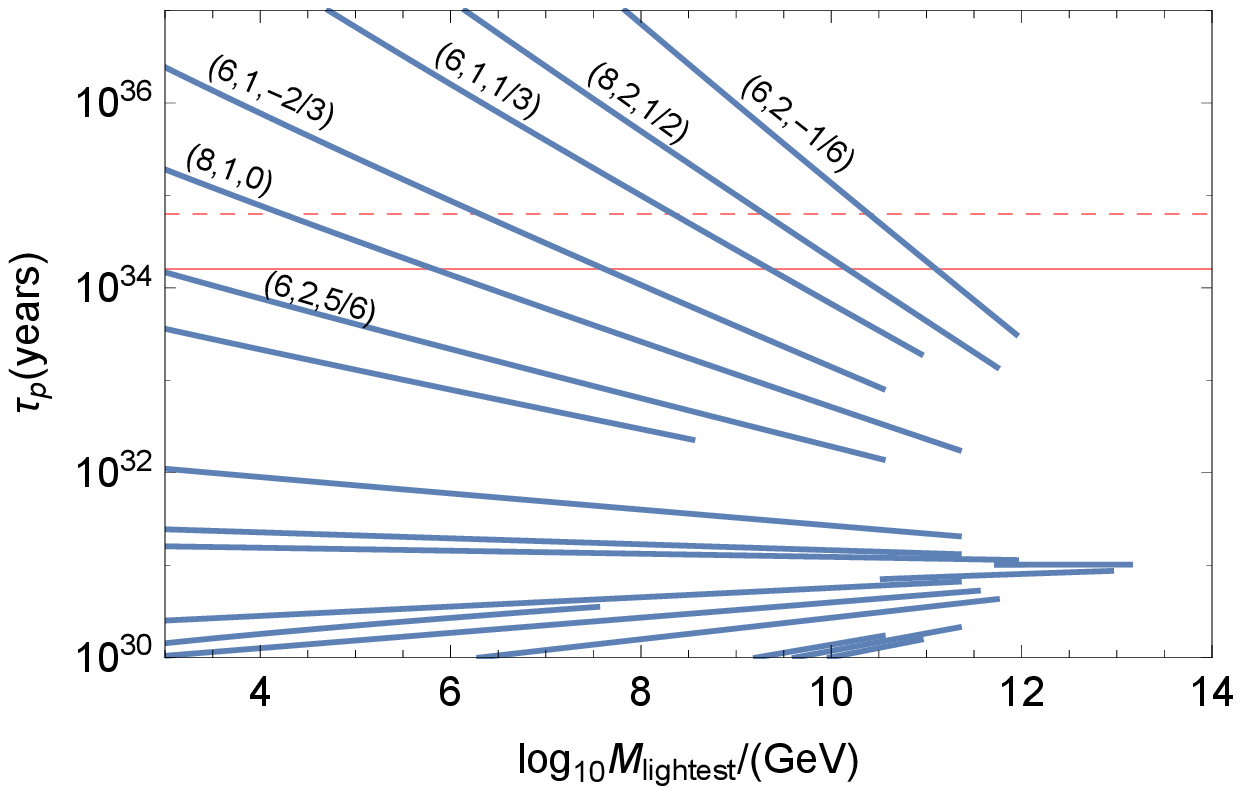}
\caption{
Proton partial lifetime, $\tau_p$, in the case when $({\bf 6},{\bf 3},1/3)$ is the \textit{second} lightest SM-decomposed multiplet.
The horizontal axis is again the mass of the leading lightest multiplet.
$\alpha_s^{(5)}(M_Z) = 0.1181+3\cdot0.0011$ (left), $0.1181-3\cdot0.0011$ (right).
The red lines show the current bound at SK (solid), and the $3\sigma$ discovery potential at HK with a 10~year exposure of 1-tank (dashed).
Each blue line with a label shows the proton partial lifetime when the \textit{leading} lightest multiplet is as indicated by the label.
The unlabelled blue lines correspond to cases with a different second lightest multiplet, and it is evident that all these cases are already excluded by SK.
Interestingly, if the leading lightest multiplet is $({\bf 8},{\bf1},1)$ or $({\bf 6},{\bf2},5/6)$, the model can evade the current bound
 and predicts a proton lifetime within the coverage of HK.
}
\label{fig3}
\end{figure}

\begin{figure}[tbp]
\includegraphics[width=8cm]{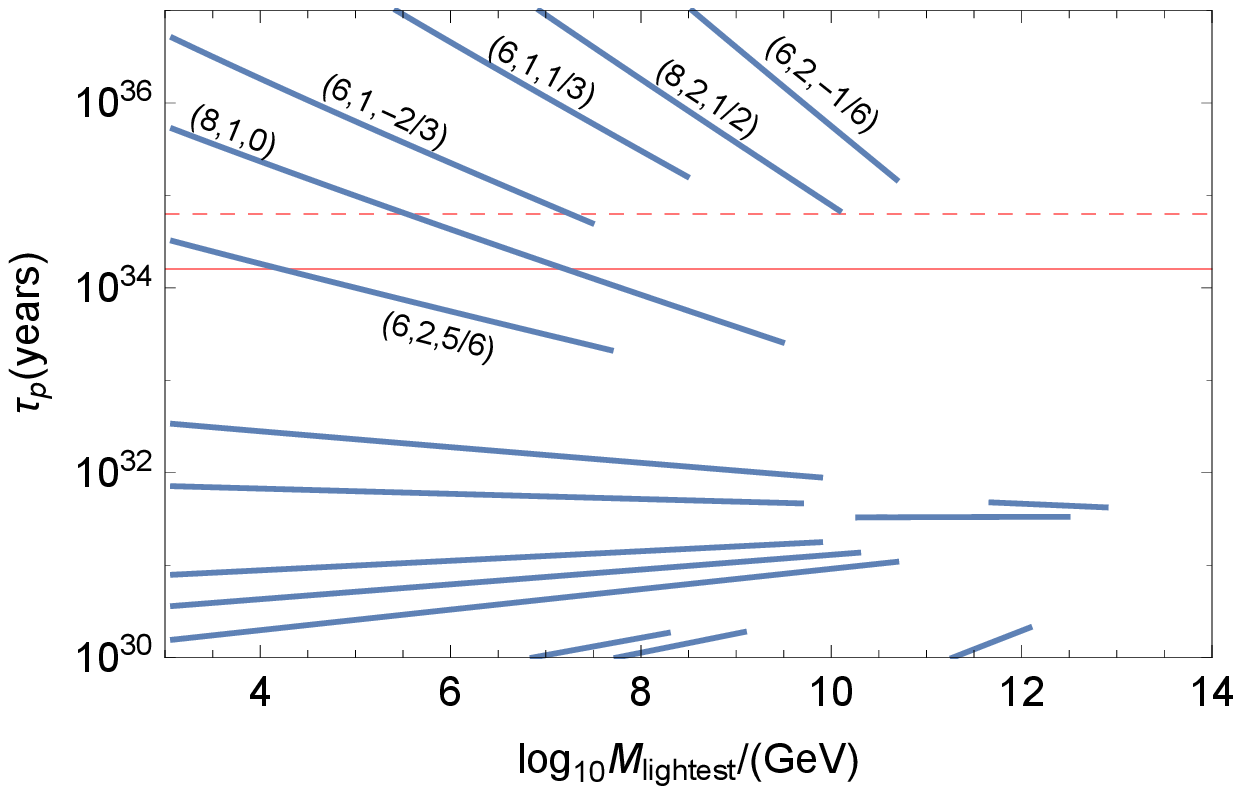}
\includegraphics[width=8cm]{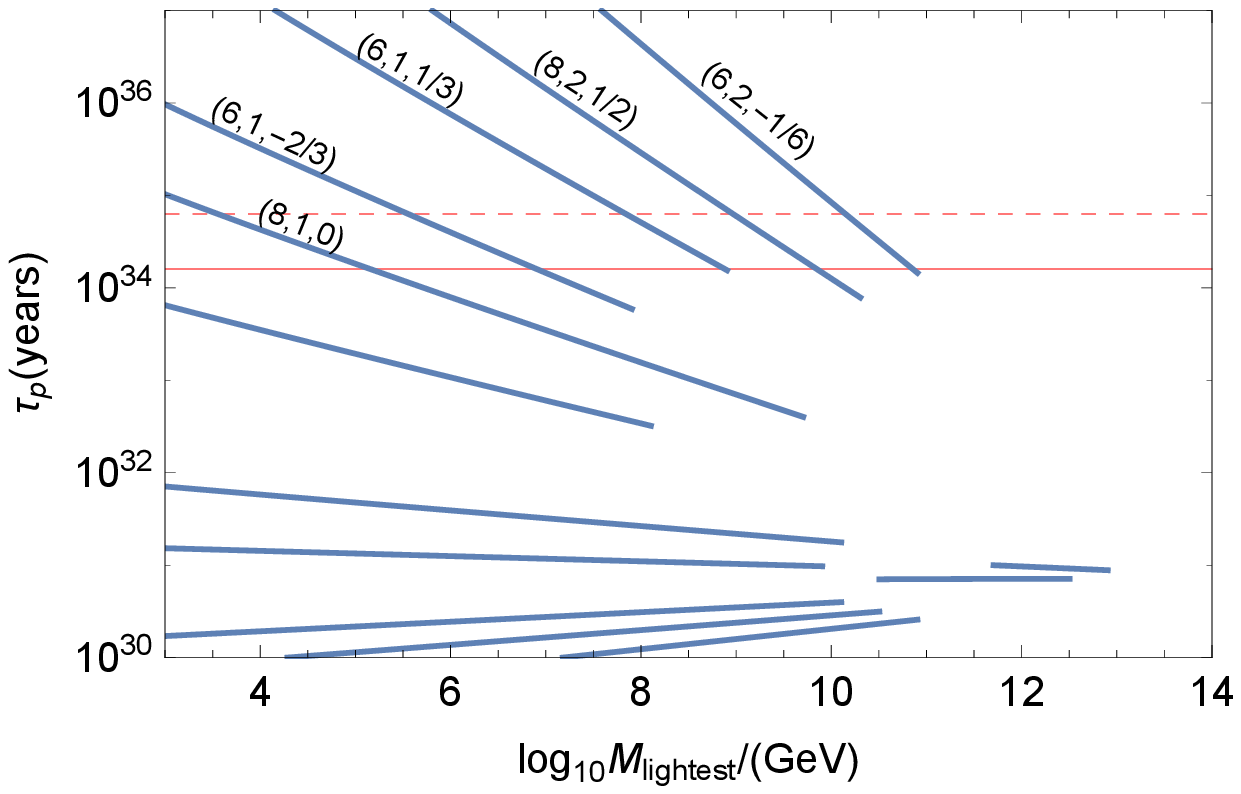}
\caption{
Proton partial lifetime, $\tau_p$, in the case when $({\bf 8},{\bf 3},0)$ is the \textit{second} lightest SM-decomposed multiplet.
The horizontal axis is again the mass of the leading lightest multiplet.
$\alpha_s^{(5)}(M_Z) = 0.1181+3\cdot0.0011$ (left), $0.1181-3\cdot0.0011$ (right).
The red lines show the current bound at SK (solid), and the $3\sigma$ discovery potential at HK with a 10~year exposure of 1-tank (dashed).
Each blue line with a label shows the proton partial lifetime when the \textit{leading} lightest multiplet is as indicated by the label.
The unlabelled blue lines correspond to cases with a different second lightest multiplet, and it is evident that all these cases are already excluded by SK.
Interestingly, if the leading lightest multiplet is $({\bf 6},{\bf2},5/6)$, the model can evade the current bound
 and predicts a proton lifetime within the coverage of HK.
}
\label{fig4}
\end{figure}

\begin{figure}[tbp]
\includegraphics[width=8cm]{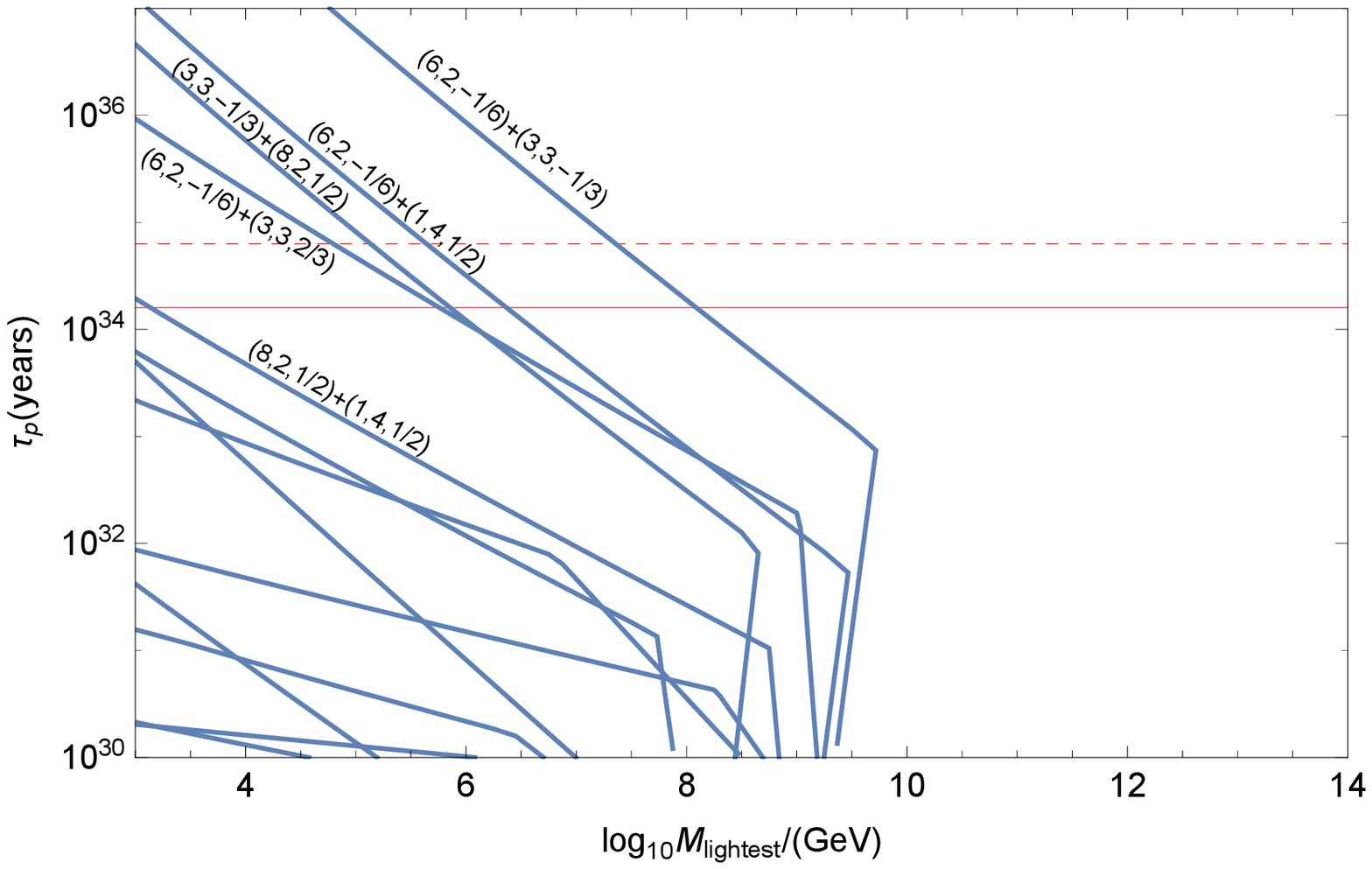}
\includegraphics[width=8cm]{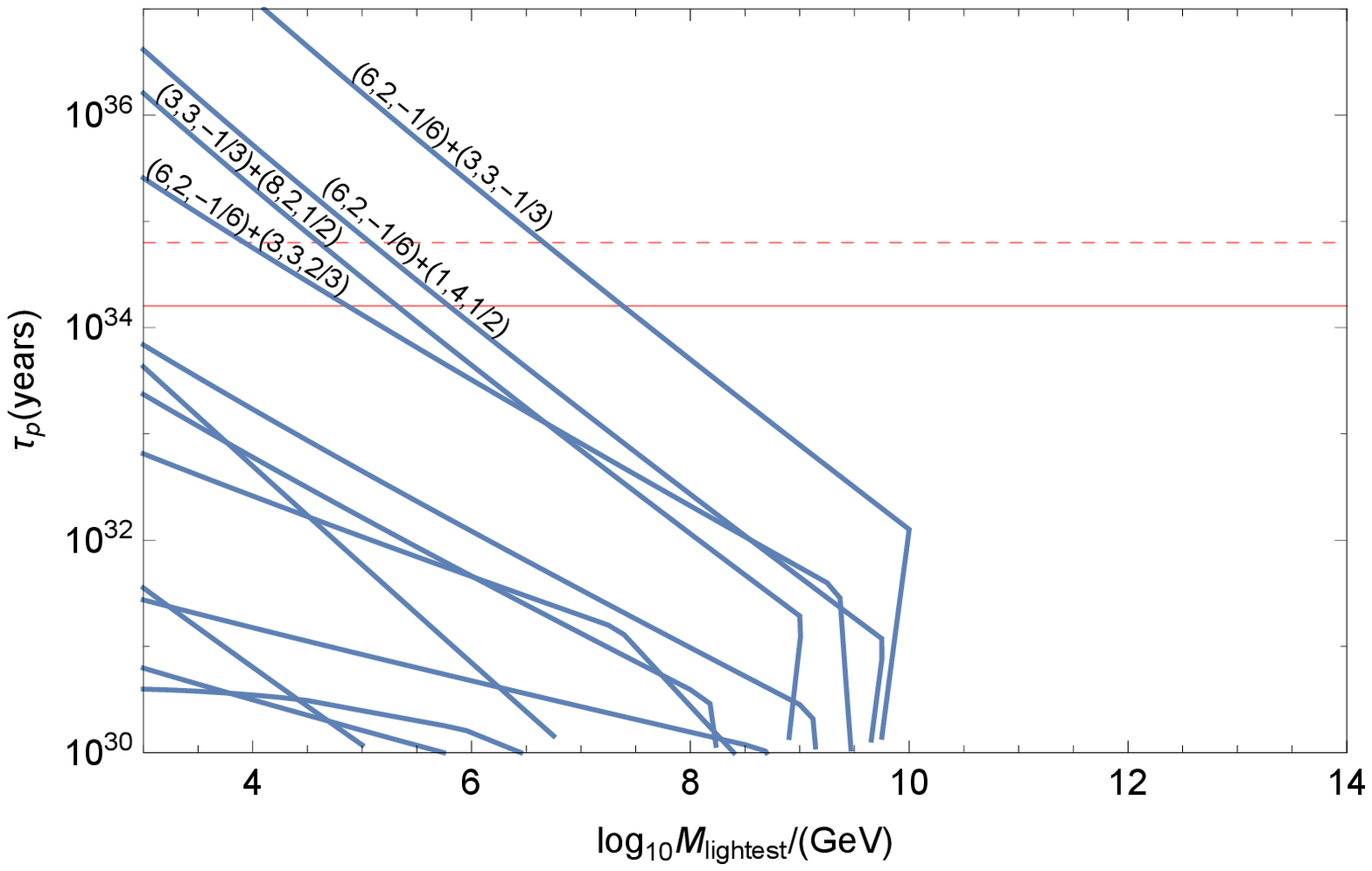}
\caption{
Proton partial lifetime, $\tau_p$, in the case when the two light SM-decomposed multiplets do \textit{not} include $({\bf 6},{\bf 3},1/3)$ or $({\bf 8},{\bf 3},0)$.
The horizontal axis is the mass of the leading lightest multiplet.
$\alpha_s^{(5)}(M_Z) = 0.1181+3\cdot0.0011$ (left), $0.1181-3\cdot0.0011$ (right).
The red lines show the current bound at SK (solid), and the $3\sigma$ discovery potential at HK with a 10~year exposure of 1-tank (dashed).
Each blue line with a label shows the proton partial lifetime when the leading and second lightest multiplets are as indicated by the label.
The unlabelled blue lines correspond to cases with a different second lightest multiplet, and it is evident that all these cases are already excluded by SK.
}
\label{fig5}
\end{figure}

In Fig.\ref{fig1},
we plot the proton partial lifetime for $p \to \pi^0 e^+$ process, $\tau_p$, when $({\bf 6},{\bf 3},1/3)$ is the \textit{leading} lightest SM-decomposed multiplet 
(except for the doublet and triplet in $\bf 5$).
We consider all possibilities for the second lightest multiplet.
However, since $({\bf 6},{\bf 3},1/3)$ is contained in $\bf 50$, we do not employ a multiplet in $\bf 70$ as the second lightest one, to maintain asymptotic freedom of the SU(5) gauge theory as mentioned in the first part of this section.
The point at which all the lines gather corresponds to the case when the second lightest multiplet is actually mass-degenerate with the GUT gauge boson, namely,
 only $({\bf 6},{\bf 3},1/3)$ is singularly light. It is clear that this case is excluded by SK.
All lines are below the current experimental bound except when the second lightest multiplet is $({\bf 10},{\bf1},-1)$ coming from $\overline{\bf 35}$.
When the second lightest multiplet is $({\bf 10},{\bf1},-1)$, the proton lifetime can be far above the HK search range,
 which signals that it is impossible to put a phenomenologically meaningful bound on the proton lifetime.

Fig.\ref{fig2} is analogous to Fig.\ref{fig1}, except that $({\bf 8},{\bf 3},0)$ is the leading lightest multiplet.
The point at which all the lines gather corresponds to the case when only $({\bf 8},{\bf 3},0)$ is singularly light. It is clear that this case is excluded by SK.
If $\alpha_s^{(5)}(M_Z)$ is larger than the current central value, six possibilities for the second lightest multiplet remain viable despite the SK bound.
Interestingly, four out of the six cases are nearly totally covered by HK, and hence our attempt to put a phenomenologically meaningful proton lifetime bound is successful in these cases.

In Figs.\ref{fig3} and \ref{fig4},
we show the proton lifetime in the scenarios in which $({\bf 6},{\bf 3},1/3)$ and 
$({\bf 8},{\bf3},0)$ are the \textit{second} lightest multiplet, respectively.
If $\alpha_s^{(5)}(M_Z)$ is larger than the current central value, the following combinations evade the SK bound and are totally covered by HK, 
 which are again considered as successful examples of our attempt.
\bea
&&{\rm \{leading \ lightest, \ second \ lightest\}} \ = \ 
\nn\\
&&\{({\bf 8},{\bf1},1),~({\bf 6},{\bf 3},1/3)\}, \ \ \ \ \{({\bf 6},{\bf2},5/6),~({\bf 6},{\bf 3},1/3)\}, \ \ \ \ \{({\bf 6},{\bf2},5/6),~({\bf 8},{\bf3},0)\}.
\nn
\eea

In Fig.\ref{fig5}, we show the proton lifetime in the scenarios
 where both $({\bf 6},{\bf 3},1/3)$ and $({\bf 8},{\bf3},0)$ are not singularly light.
In this case, either $({\bf6},{\bf 2},-1/6)$ or $({\bf8},{\bf 2},1/2)$ constitutes the two lightest multiplets
\footnote{
We note that $({\bf3},{\bf 3},-1/3)+({\bf15},{\bf 1},-1/3)$ is also possible.
However, $({\bf3},{\bf 3},-1/3)$ has mass around $10^6$ GeV for this choice,
which can cause a too rapid proton decay via $({\bf3},{\bf 3},-1/3)$ exchange.
Thus we discard this combination.
}.
\\

It is interesting to note that 
 there is a theoretical motivation to employ $\bf 45$ representation
 to build SU(5) GUT models,
 since models with ${\bf24}$, ${\bf45}$ and ${\bf5}$ scalars can accommodate realistic renormalizable Yukawa couplings.
In the model that exclusively contains ${\bf24}$, ${\bf45}$ and ${\bf5}$ scalars,
 the only case that is not excluded by SK is the one where 
 $({\bf 8},{\bf 2},1/2)$ and $({\bf 3},{\bf 3},-1/3)$ are the singularly light multiplets,
 and the proton lifetime is given by the line of
 `$({\bf 8},{\bf 2},1/2)+({\bf 3},{\bf 3},-1/3)$' in Fig.~\ref{fig5}.

The representation $\bf 40$ is not motivated in SU(5) GUT, but it is contained in 
$\bf 210$ and $\bf 144$ that break SO(10) symmetry, and so $\bf 40$ can be motivated 
 if we consider larger GUTs.
As far as we know,  there is no motivation to employ $\bf 35$ and $\bf 70$ representations,
but we note that the combination $({\bf 8},{\bf3},0)$+$({\bf 15},{\bf1},-1/3)$
provides the largest proton lifetime under our criteria that the SU(5) gauge theory must be
asymptotic free.
In this case, the GUT gauge boson mass can be as large as the Planck scale.
\\

\section{Summary}

We have surveyed all non-SUSY SU(5) models
  under the restrictions that (i) only one or two SM-decomposed multiplets (except for the electroweak and colored Higgses) are singularly light
  and that (ii) the SU(5) gauge theory is asymptotically free, namely, SU(5) representations with a large Dynkin index do not enter,
  and have attempted to derive a phenomenologically meaningful upper bound on the proton lifetime for various choices of the singularly light multiplets.

We have formulated criteria for singularly light multiplets that enhance the proton lifetime.
When two multiplets are light, the criteria are summarized as, (1) $l_A^{\rm eff}<0$, $l_B^{\rm eff}>0$, (2) sufficiently large $-l_A^{\rm eff}$ and $l_B^{\rm eff}$,
 (3)$-l_B^{\rm eff}/l_A^{\rm eff}\gtrsim1.1$.
Here, $l_A^{\rm eff}$ and $l_B^{\rm eff}$ are defined in Eqs.~(\ref{leff1}),(\ref{leff2}) and can be computed using Tables~\ref{Tab1},\ref{Tab2}.

It has been shown that when only one SM-decomposed multiplet is singularly light, the proton lifetime is always below the current SK bound.

When two multiplets are singularly light, the SK bound is evaded in a few cases, and we have successfully derived a testable upper bound in some of them.
The results are summarized as follows: 
When the leading lightest multiplet is $({\bf 6},{\bf 3},1/3)$, all cases are excluded by SK,
 except for one case where the second lightest multiplet $({\bf 10},{\bf 1},-1)$,
 in which case no meaningful bound on the proton lifetime is obtained.
When the leading lightest multiplet is $({\bf 8},{\bf 3},0)$,
 six cases possibly evade the SK bound (if $\alpha_s^{(5)}(M_Z)$ is large),
 four of which lead to a proton lifetime limited within the coverage of HK, i.e. a testable proton lifetime is obtained.
When the second lightest multiplet is $({\bf 6},{\bf 3},1/3)$,
 seven cases can evade the SK bound (if $\alpha_s^{(5)}(M_Z)$ is large), 
 two of which lead to a proton lifetime limited within the coverage of HK.
When the second lightest multiplet is $({\bf 8},{\bf 3},0)$,
 six cases can evade the SK bound (if $\alpha_s^{(5)}(M_Z)$ is large), 
 one of which leads to a proton lifetime limited within the coverage of HK.
Finally, when both the leading and second lightest multiplets differ from
 $({\bf 6},{\bf 3},1/3)$ and $({\bf 8},{\bf 3},0)$, four cases survive despite the SK bound and all of them can lead to a proton lifetime far above the HK range.

It is straightforward to restrict our study to theoretically well-motivated models, such as the model containing only ${\bf24}$+${\bf45}$+${\bf5}$ scalars, which can give realistic renormalizable Yukawa couplings.

\appendix

\section*{Appendix: Gauge coupling unification}

The gauge coupling unification conditions in SUSY SU(5) and SO(10) GUTs \cite{Hisano:1992jj} are written as\footnote{
The SUSY particle contributions in the equations are more precisely written as
 \begin{eqnarray}
-2 \ln \frac{m_{\rm SUSY}}{m_Z}
&=& -\frac85 \ln \frac{\mu_H}{m_Z} - \frac25 \ln \frac{m_H}{m_Z}
+ 4 \ln \frac{M_3}{M_2} 
+ \frac35 \ln \frac{m_{\tilde q^c}^3 m_{\tilde d^c}^2 m_{\tilde e^c}}{m_{\tilde q}^4 m_{\tilde \ell}^2}, 
\\
8 \ln \frac{m_{\rm SUSY}}{m_Z}
&=&
 4 \ln \frac{M_3 M_2}{m_Z^2}
+ 3 \ln \frac{m_{\tilde q}^2}{m_{\tilde u} m_{\tilde e^c}},
\end{eqnarray}
where $\mu_H$, $m_H$ are higgsino and heavier Higgs masses,
$M_3$ and $M_2$ are gluino and wino masses.
}
\begin{eqnarray}
&&-\frac{2}{\alpha_3(m_Z)} + \frac{3}{\alpha_2(m_Z)} - \frac{1}{\alpha_1(m_Z)}
= \frac1{2\pi} \left(\frac{12}5 \ln \frac{M_{H}}{m_Z} -2 \ln \frac{m_{\rm SUSY}}{m_Z}
\right),
\label{susyHc}\\
&&-\frac{2}{\alpha_3(m_Z)} - \frac{3}{\alpha_2(m_Z)} + \frac{5}{\alpha_1(m_Z)}
= \frac1{2\pi} \left(36 \ln \frac{M_U}{m_Z} + 8 \ln \frac{m_{\rm SUSY}}{m_Z}
\right),
\label{susyMG}
\end{eqnarray}
where\footnote{
The suffix $G$ stands for the multiplets of GUT gauge bosons.
In SU(5) GUT, the GUT gauge boson is just $({\bf3},{\bf2},-5/6)+c.c.$
For convenience, we also show the contribution from the GUT gauge bosons in SO(10) GUT.
In SU(5) GUT, just eliminate the other gauge boson, or 
impose the SU(5) condition, $M_{G({\bf3},{\bf2},1/6)}= M_{G({\bf3},{\bf1},2/3)}=M_{G({\bf1},{\bf1},1)}$.
}
\begin{eqnarray}
M_H &\equiv& \frac{M_{G({\bf3},{\bf2},1/6)}^4}{M_{G({\bf1},{\bf1},1)} M_{G({\bf3},{\bf1},2/3)}^3 } M_{H_C} \prod_i M_i^{l_A^i},
\\
M_U^{36} &\equiv& \frac{M_{G({\bf3},{\bf2},5/6)}^{24} M_{G({\bf1},{\bf1},1)}^{12} M_{G({\bf3},{\bf1},2/3)}^{12}}{M_{G({\bf3},{\bf2},1/6)}^{24}} \prod_i M_i^{l_B^i},
\end{eqnarray}
and $M_{H_C}$ stands for the lightest colored Higgs mass, and $i$ runs over all SM-decomposed multiplets other than the would-be-Nambu-Goldstone modes.
We define
\begin{equation}
l_A = \frac{5}{12} \Delta(2 l_3 - 3 l_2 +  l_1),
\qquad l_B = \frac16 \Delta(2 l_3 + 3 l_2 - 5 l_1), \label{lAB}
\end{equation}
for each multiplet $i$ under SM gauge symmetry, which are listed in Tables \ref{Tab1} and \ref{Tab2}.
We note that the contribution to the beta coefficients of the vector-like multiplet
 is given as 
$\Delta b_i^{\rm SUSY} = l_i$.
Since the would-be-Nambu-Goldstone modes eaten by the GUT gauge bosons
are lacking in the SU(5) multiplets, 
we obtain 
 \begin{equation}
\sum_i l_A^i = 0, \qquad
\sum_i l_B^i = 2.
\end{equation}
Also,
\begin{equation}
M_H^{\rm SUSY} \simeq 10^{16} \ {\rm GeV}, \qquad
M_U^{\rm SUSY} \simeq 2 \times 10^{16} \ {\rm GeV}.
\end{equation}

The gauge coupling unification conditions in non-SUSY case is written as\footnote{
We remark that
the degree $i$ in the formula corresponds to a vector-like set of (not self-conjugate) multiplets in SUSY case.
The degree of scalar in non-SUSY case is counted without the pair 
and we treat as a real scalar for self-conjugate reps. such as $({\bf8},{\bf1},0)$ and (${\bf1},{\bf3},0$).
 }
\begin{eqnarray}
&&-\frac{2}{\alpha_3(m_Z)} + \frac{3}{\alpha_2(m_Z)} - \frac{1}{\alpha_1(m_Z)}
= \frac1{2\pi} 
\frac{2}5 \ln \frac{M_{H}}{m_Z} ,
\label{nonsusyHc}\\
&&-\frac{2}{\alpha_3(m_Z)} - \frac{3}{\alpha_2(m_Z)} + \frac{5}{\alpha_1(m_Z)}
= \frac1{2\pi} 
44 \ln \frac{M_U}{m_Z} .
\label{nonsusyMG}
\end{eqnarray}
\begin{eqnarray}
M_H &=& \frac{M_{G({\bf3},{\bf2},1/6)}^{42}}{M_{G({\bf1},{\bf1},1)}^{21/2} M_{G({\bf3},{\bf1},2/3)}^{63/2} } M_{H_C} \prod_i M_i^{l_A^i},
\label{hcappendix}\\
M_U^{44} &=& \frac{M_{G({\bf3},{\bf2},5/6)}^{42} M_{G({\bf1},{\bf1},1)}^{21} M_{G({\bf3},{\bf1},2/3)}^{21}}{M_{G({\bf3},{\bf2},1/6)}^{42}} \prod_i M_i^{l_B^i}.
\label{uappendix}
\end{eqnarray}
The contribution to a beta coefficient from a complex scalar in representation $R$ is $\frac{1}{3}C(R)$,
 while that from a pair of chiral supermultiplets is $2\left(\frac{1}{3}+\frac{2}{3}\right)C(R)$.
Their ratio, $\frac{1}{6}$, accounts for the ratio of the coefficients of $\log (M_H/m_Z)$ in Eq.~(\ref{susyHc}) and (\ref{nonsusyHc}).
The contribution to a beta coefficient from a massive gauge boson in representation $R$ is $\left(-\frac{11}{3}+\frac{1}{6}\right)C(R)$
 (the factor $+\frac{1}{6}$ comes from a real scalar eaten by the gauge boson),
 while that from a massive gauge supermultiplet is $(-3+1)C(R)$
 (the factor $+1$ comes from a chiral supermultiplet eaten by the gauge supermultiplet).
Their ratio, $\frac{7}{4}$, accounts for the ratio of the coefficients of $\log (M_G/m_Z)$ in Eq.~(\ref{susyMG}) and (\ref{nonsusyMG}).
From Eqs.~(\ref{hcappendix}),(\ref{uappendix}), we obtain\footnote{
For $\alpha_s^{(5)}(M_Z) = 0.1181+3 \cdot 0.0011$,
\begin{equation}
M_H^{\rm non-SUSY} \simeq 5.96 \times 10^{90} \ {\rm GeV}, \qquad
M_U^{\rm non-SUSY} \simeq 6.04 \times 10^{13} \ {\rm GeV}.
\end{equation}
For $\alpha_s^{(5)}(M_Z) = 0.1181-3 \cdot 0.0011$,
\begin{equation}
M_H^{\rm non-SUSY} \simeq 2.71 \times 10^{84} \ {\rm GeV}, \qquad
M_U^{\rm non-SUSY} \simeq 5.24 \times 10^{13} \ {\rm GeV}.
\end{equation}
}
\begin{equation}
M_H^{\rm non-SUSY} \simeq 10^{87} \ {\rm GeV}, \qquad
M_U^{\rm non-SUSY} \simeq 5 \times 10^{13} \ {\rm GeV},
\end{equation}
at one-loop level. 
Obviously, $M_H$ is too large and $M_U$ is too small to have the colored Higgs mass $M_{H_C}$ in a reasonable range (below the Planck scale) and to evade the SK bound on the dimension-six proton decay.
To remedy this, we assume either that
(1) a multiplet(s) with $l_A <0$ and $l_B >0$ is singularly light (which is the case studied in this paper),
or that
(2)
$M_{G({\bf1},{\bf1},1)}$ is much smaller than $M_{G({\bf3},{\bf2},1/6)}$ and $M_{G({\bf3},{\bf1},2/3)}$ in SO(10),
namely, $SU(2)_R$ symmetry remains at an intermediate scale (see, e.g., Ref.~\cite{Babu:2015bna}).


\section*{Acknowledgement}
The authors thank Takaaki Kajita for motivating this work and Tsuyoshi Nakaya for fruitful discussions.
This work is partially supported by Scientific Grants by the Ministry
of Education, Culture, Sports, Science and Technology of Japan
(Nos.~16H00871, 16H02189, 17K05415 and 18H04590).


\begin{table}[p]
\begin{center}
\begin{tabular}{cc}
\begin{tabular}{|c|c|c|c|c|c|} \hline
$\bf5$ & $l_3$ & $l_2$ & $l_1$ & $l_A$ &
 $l_B$ \\ \hline
$({\bf 1},{\bf 2},\frac12) $ & 0 & 1 & $\frac35$ & $-1$ & $0$ \\ \hline
$({\bf 3},{\bf 1},-\frac13) $ & 1 & 0 & $\frac25$ & $1$ & $0$ \\ \hline \hline
total & 1&1 &1 &0 &0  \\ \hline
\end{tabular}
&
\begin{tabular}{|c|c|c|c|c|c|} \hline
$\bf10$ & $l_3$ & $l_2$ & $l_1$ & $l_A$ &
 $l_B$ \\ \hline
 $({\bf 1},{\bf 1},1) $ & 0 & 0 & $\frac65$ & $\frac12$ & $-1$ \\ \hline
$(\bar{\bf 3},{\bf 1},-\frac23) $ & 1 & 0 & $\frac85$ & $\frac{3}2$ & $-1$ \\ \hline
$({\bf 3},{\bf 2},\frac16) $ & 2 & 3 & $\frac15$ & $-2$ & $2$ \\ \hline \hline
total & 3&3 &3 &0 &0  \\ \hline
\end{tabular} \\
\\
\begin{tabular}{|c|c|c|c|c|c|} \hline
$\bf15$ & $l_3$ & $l_2$ & $l_1$ & $l_A$ &
 $l_B$ \\ \hline
$({\bf 1},{\bf 3},1)$ 
   & 0 & 4 & $\frac{18}5$ & $-\frac{7}2$ & $-1$  \\ \hline
 $({\bf 3},{\bf 2},\frac16)$ 
   & 2 & 3 & $\frac15$ & $-2$ & 2  \\ \hline
$({\bf 6},{\bf 1},-\frac23)$ 
   & 5 & 0 & $\frac{16}5$ & $\frac{11}2$ & $-1$  \\ \hline \hline
total & 7&7 &7 &0 &0  \\ \hline
\end{tabular}
&
\begin{tabular}{|c|c|c|c|c|c|} \hline
$\bf24$ & $l_3$ & $l_2$ & $l_1$ & $l_A$ &
 $l_B$ \\ \hline
 $({\bf 1},{\bf 1},0)$ 
   & 0 & 0 & $0$ & $0$ & 0 \\ \hline
 $({\bf 1},{\bf 3},0)$ 
   & 0 & 2 & $0$ & $-\frac52$ & $1$  \\ \hline
 $({\bf 3},{\bf 2},-\frac56)$ 
   & 2 & 3 & $5$ & $0$ & $-2$ \\ \hline
 $({\bf 8},{\bf 1},0)$ 
   & 3 & 0 & $0$ & $\frac52$ & $1$  \\ \hline  \hline
total & 5&5 &5 &0 &0  \\ \hline
%
\end{tabular} \\
\\
\begin{tabular}{|c|c|c|c|c|c|} \hline
$\bf35$ & $l_3$ & $l_2$ & $l_1$ & $l_A$ &
 $l_B$ \\ \hline
 $({\bf 1},{\bf 4},-\frac32) $ 
   & 0 & 10 & $\frac{54}5$ & $-8$ & $-4$  \\ \hline
 $(\bar{\bf 3},{\bf 3},-\frac23)$ 
   & 3 & 12 & $\frac{24}5$ & $-\frac{21}2$ & 3  \\ \hline
  $(\bar{\bf 6},{\bf 2},\frac16)$ 
   & 10 & 6 & $\frac25$ & $1$ & $6$ \\ \hline  
  $(\overline{\bf 10},{\bf 1},1) $ 
   & 15 & 0 & $12$ & $\frac{35}2$ & $-5$  \\ \hline   \hline
total & 28&28 &28 &0 &0  \\ \hline 
\end{tabular}
&
\begin{tabular}{|c|c|c|c|c|c|} \hline
$\bf40$ & $l_3$ & $l_2$ & $l_1$ & $l_A$ &
 $l_B$ \\ \hline
 $({\bf 1},{\bf 2},-\frac32)$ 
   & 0 & 1 & $\frac{27}5$ & $1$ & $-4$  \\ \hline
 $({\bf 3},{\bf 2},\frac16)$ 
   & 2 & 3 & $\frac15$ & $-2$ & 2  \\ \hline
  $(\bar{\bf 3},{\bf 1},-\frac23)$ 
   & 1 & 0 & $\frac85$ & $\frac32$ & $-1$ \\ \hline  
  $(\bar{\bf 3},{\bf 3},-\frac23)$ 
   & 3 & 12 & $\frac{24}5$ & $-\frac{21}2$ & $3$  \\ \hline    
   $({\bf 8},{\bf 1},1) $ 
   & 6 & 0 & $\frac{48}5$ & $9$ & $-6$\\ \hline
   $(\bar{\bf 6},{\bf 2},\frac16)$
   & 10 & 6 & $\frac{2}5$ & $1$ & $6$ \\ \hline \hline
total & 22&22 &22 &0 &0  \\ \hline
\end{tabular}
\end{tabular}
\end{center}

\caption{The list of the decomposed multiplets under $SU(3)_c \times SU(2)_L \times U(1)_Y$ in the respective SU(5)
representations,
and the contributions to the beta coefficients $\Delta b_i= l_i/6$.
For $\bf 24$ representation, the contribution of a real scalar to the beta function is shown.
$l_{A,B}$ are defined in Eq.(\ref{lAB}) and they are used to specify the contribution to the gauge coupling unification
conditions.}
\label{Tab1}
\end{table}

\begin{table}[p]
\begin{center}
\begin{tabular}{cc}
\begin{tabular}{|c|c|c|c|c|c|} \hline
$\bf45$ & $l_3$ & $l_2$ & $l_1$ & $l_A$ &
 $l_B$ \\ \hline
 $({\bf 1},{\bf 2},\frac12)$ 
   & 0 & 1 & $\frac35$ & $-1$ & 0  \\ \hline
 $({\bf 3},{\bf 1},-\frac13)$ 
   & 1 & 0 & $\frac25$ & $1$ & 0  \\ \hline
   $({\bf 3},{\bf 3},-\frac13) $ 
   & 3 & 12 & $\frac{6}5$ & $-12$ & $6$ \\ \hline
   $(\bar{\bf 3},{\bf 1},\frac43) $ 
   & 1 & 0 & $\frac{32}5$ & $\frac{7}2$ & $-5$   \\ \hline
   $(\bar{\bf 3},{\bf 2},-\frac76) $ 
   & 2 & 3 & $\frac{49}5$ & $2$ & $-6$  \\ \hline
   $(\bar{\bf 6},{\bf 1},-\frac13) $ 
   & 5 & 0 & $\frac{4}5$ & $\frac{9}2$ & $1$   \\ \hline
  $({\bf 8},{\bf 2},\frac12) $ 
   & 12 & 8 & $\frac{24}5$ & $2$ & $4$  \\ \hline \hline
total & 24&24 &24 &0 &0  \\ \hline
\end{tabular}
&
\begin{tabular}{|c|c|c|c|c|c|} \hline
$\bf50$ & $l_3$ & $l_2$ & $l_1$ & $l_A$ &
 $l_B$ \\ \hline
 $({\bf 1},{\bf 1},-2)$ 
   & 0 &0 & $\frac{24}5$ & $2$ & $-4$  \\ \hline
 $({\bf 3},{\bf 1},-\frac13)$ 
   & 1 & 0 & $\frac25$ & $1$ & 0  \\ \hline
   $(\bar{\bf 3},{\bf 2},-\frac76) $ 
   & 2 & 3 & $\frac{49}5$ & $2$ & $-6$ \\ \hline
   $(\bar{\bf 6},{\bf 3},-\frac13) $ 
   & 15 & 24 & $\frac{12}5$ & $-\frac{33}2$ & $15$   \\ \hline
   $({\bf 6},{\bf 1},\frac43) $ 
   & 5 & 0 & $\frac{64}5$ & $\frac{19}2$ & $-9$  \\ \hline
   $({\bf 8},{\bf 2},\frac12) $ 
   & 12 & 8 & $\frac{24}5$ & $2$ & $4$  \\ \hline \hline
total & 35&35 &35 &0 &0  \\ \hline
\end{tabular} \\
\\
\begin{tabular}{|c|c|c|c|c|c|} \hline
$\bf70$ & $l_3$ & $l_2$ & $l_1$ & $l_A$ &
 $l_B$ \\ \hline
 $({\bf 1},{\bf 2},\frac12)$ 
   & 0 &1 & $\frac{3}5$ & $-1$ & $0$  \\ \hline
 $({\bf 3},{\bf 1},-\frac13)$ 
   & 1 & 0 & $\frac25$ & $1$ & 0  \\ \hline
 $({\bf 1},{\bf 4},\frac12)$ 
   & 0 &10 & $\frac{6}5$ & $-12$ & $4$  \\ \hline
   $({\bf 3},{\bf 3},-\frac13) $ 
   & 3 & 12 & $\frac{6}5$ & $-12$ & $6$ \\ \hline
   $(\bar{\bf 3},{\bf 3},\frac43) $ 
   & 3 & 12 & $\frac{96}5$ & $-\frac{9}2$ & $-9$   \\ \hline
   $({\bf 6},{\bf 2},-\frac76) $ 
   & 10 & 6 & $\frac{98}5$ & $9$ & $-10$  \\ \hline
   $({\bf 8},{\bf 2},\frac12) $ 
   & 12 & 8 & $\frac{24}5$ & $2$ & $4$  \\ \hline
      $({\bf 15},{\bf 1},-\frac13) $ 
   & 20 & 0 & $2$ & $\frac{35}2$ & $5$  \\ \hline \hline
total & 49&49 &49 &0 &0  \\ \hline
\end{tabular}
&
\begin{tabular}{|c|c|c|c|c|c|} \hline
$\bf75$ & $l_3$ & $l_2$ & $l_1$ & $l_A$ &
 $l_B$ \\ \hline
 $({\bf 1},{\bf 1},0)$ 
   & 0 &0 & $0$ & $0$ & $0$  \\ \hline
$({\bf 3},{\bf 1},\frac53) $ 
   & 1 & 0 & $10$ & $5$ & $-8$  \\ \hline
 $({\bf 3},{\bf 2},-\frac56)$ 
   & 2 & 3 & $5$ & $0$ & $-2$ \\ \hline
$({\bf 6},{\bf 2},\frac56) $ 
   & 10 & 6 & $10$ & $5$ & $-2$  \\ \hline
 $({\bf 8},{\bf 1},0)$ 
   & 3 & 0 & $0$ & $\frac52$ & $1$  \\ \hline  
$({\bf 8},{\bf 3},0)$ 
   & 9 & 16 & $0$ & $-\frac{25}{2}$ & $11$   \\ \hline \hline
total & 25&25 &25 &0 &0  \\ \hline
\end{tabular}
\end{tabular}
\end{center}

\caption{Continuation of the list in Table \ref{Tab1}.
For $\bf 75$ representation, the contribution of a real scalar to the beta function is shown.
Note that $({\bf3},{\bf1},-4/3)$ and $({\bf3},{\bf3},-1/3)$ in {\bf 45} can cause proton decay 
in addition to the usual colored Higgs $({\bf3},{\bf1},-1/3)$ in ${\bf 5}$ and $\bf 50$.
}
\label{Tab2}
\end{table}

\end{document}